\documentclass[twocolumn,prb,showpacs,preprintnumbers,amsmath,amssymb]{revtex4}

\usepackage{graphicx}
\usepackage{dcolumn}
\usepackage{bm}
\begin{document}

\title{Low-energy excitations in the magnetized state of the bond-alternating quantum $S=1$ chain system NTENP.}

\author{L. P. Regnault}
\affiliation{CEA-Grenoble, DRFMC-SPSMS-MDN, 17 rue des Martyrs,
38054 Grenoble Cedex 9, France.}

\author{A. Zheludev}
\affiliation{Condensed Matter Sciences Division, Oak Ridge
National Laboratory, Oak Ridge, Tennessee 37831-6393, USA.}

\author{M. Hagiwara}
\affiliation{KYOKUGEN, Osaka University, 1-3 Machikaneyama,
Toyonaka 560-8531, Japan.}

\author{A. Stunault}
\affiliation{Institut Laue Langevin, 6 rue J. Horowitz, 38042
Grenoble Cedex 9, France.}

\date{\today}

\begin{abstract}
High intensity inelastic neutron scattering experiments on the
$S$=1 quasi-one-dimensional bond-alternating antiferromagnet
Ni(C$_9$D$_{24}$N$_4$)(NO$_2$)ClO$_4$ are performed in magnetic
fields of up to 14.8~T. Excitation in the high field magnetized
quantum spin solid (ordered) phase are investigated. In addition
to the previously observed coherent long-lived gap excitation [M.
Hagiwara {\it et al.}, Phys. Rev. Lett {\bf 94}, 177202 (2005)], a
broad continuum is detected at lower energies. This observation is
consistent with recent numerical studies, and helps explain the
suppression of the lowest-energy gap mode in the magnetized state
of NTENP. Yet another new feature of the excitation spectrum is
found at slightly higher energies, and appears to be some kind of
multi-magnon state.
\end{abstract}

\maketitle

\section{Introduction}
The spin dynamics of gapped one-dimensional (1-D) quantum magnets
in high magnetic fields has recently become an area of intensive
research. The lowest-energy excitations in these systems are
typically an $S=1$ triplet, and therefore subject to Zeeman
splitting by an external field. With increasing field the gap in
one of the three modes decreases and eventually vanishes, leading
to a 1-D Bose condensation of
magnons.\cite{Schulz86,Katsumata89,Affleck90,Takahashi91} A
qualitatively new state emerges at higher fields. In the presence
of magnetic anisotropy it is characterized by long-range Neel
order. Despite that, its dynamical properties are nothing like
those of conventional 3-D magnets. Quasiclassical spin wave theory
(SWT) breaks down on the qualitative level, and the ground states
are exotic ``quantum spin solids''. Such is the case for the
Haldane gap compounds
NDMAP~\cite{Honda98,Chen2001,Zheludev2002,Zheludev2003,Zheludev2005rev}
 and NENP,\cite{Regnault} where at
$H>H_c$ there are {\it three} distinct magnon branches, while SWT
can account for only {\it two}.

The excitation spectrum in the high-field phase is non-universal,
but can qualitatively vary from one spin system to another. In a
recent work we investigated the bond-alternating dimerized $S=1$
chain compound NTENP.\cite{Narumi2001,Zheludev2004-2,Regnault2004}
Like the structurally similar  NDMAP, at $H=0$ this material has a
singlet ground state and a triplet of coherent gap excitations. It
goes through a soft mode antiferromagnetic ordering transition at
$H_c\approx 9.7$~T in a field applied parallel to the spin chains,
and $H_c\approx 13$~T in a transverse field.~\cite{Narumi2001}
However, its dynamical properties in a field are markedly
different: only {\it one} coherent mode survives in the high-field
phase of NTENP.\cite{Hagiwara2005,Zheludev2005rev} The
highest-energy gap mode disappears well below $H_c$. This is due
to interactions between magnons that mix single-magnon and
two-magnon states, opening an effective decay channel for the
upper mode.\cite{Suzuki2005,Zhitomirsky2005,Kolezhuk2005} In
addition, the lowest-energy mode (the one that goes soft), is a
sharp long-lived excitation at all fields below the transition,
but is absent in the magnetized phase. Thus, only the middle mode
persists above $H_c$. This one-mode picture is at present poorly
understood. In the present paper we report a high intensity
inelastic neutron scattering study that sheds new light on this
peculiar behavior. At high fields, apart from the previously seen
single coherent mode, we find a novel gapped low-energy excitation
continuum.

\section{Experimental procedures and results}

The structural elements of NTENP are discussed elsewhere, so only
the key features are summarized here. NTENP crystallizes in the
triclinic system (space group $P$$\bar{1}$) with lattice constants
$a$=10.747(1)$\AA$, $b$=9.413(2)$\AA$, $c$=8.789(2)$\AA$,
$\alpha$=95.52(2)$^{\circ}$, $\beta$=108.98(3)$^{\circ}$ and
$\gamma$=106.83(3)$^{\circ}$.  The Ni$^{2+}$ ions are bridged by
nitrito groups along the $a$ axis with alternating bond distances
of 2.142(3) and 2.432(6) $\AA$, and corresponding exchange
constants are estimated as $J_1=2.1$~meV and $J_2=4.7$~meV.
Single-ion easy-plane anistropy of type $DS_z^2$ is significant
with $D/\overline{J}=0.25$, where $\overline{J}=(J_{1}+J_{2})/2$
is the average coupling constant. In the experiments we employed a
fully deuterated NTENP single crystal of approximate volume
$10\times 8\times 4$~mm$^{3}$. Sample environment was a 15~T split
coil superconducting magnet with a dilution refrigerator insert.
Inelastic neutron data were collected at the IN14 cold-neutron
spectrometer at Institut Laue-Langevin. In most cases neutrons
with a fixed final energy (wave number) of 2.74~meV (1.15
\AA$^{-1}$) were used with a Be higher-order filter positioned
after the sample and a horizontally focusing Pyrolitic Graphite
(PG) analyzer. No additional beam collimation devices were used.
The sample was in all cases mounted with the $b$ axis vertical
(parallel to the applied field), allowing access to wave vectors
in the $(h,0,l)$ reciprocal-space plane.
\begin{figure}
\includegraphics[width=3.4in]{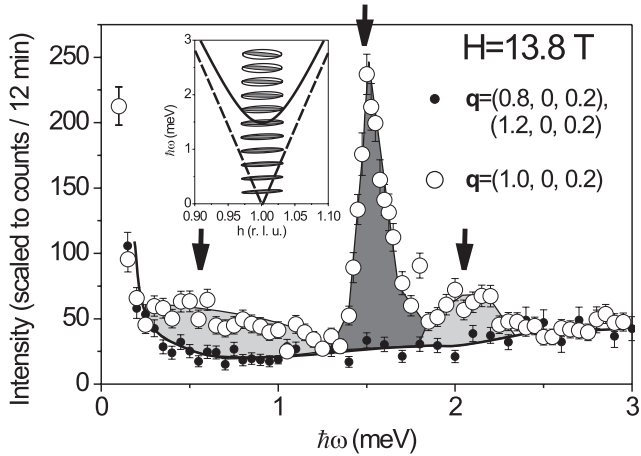}
\caption{A typical inelastic scan meaured at the 1D AF zone-center
in NTENP at $H=13.8$~T$>H_c$ (open circles). The background was
measured away from the zone-center position (solid circles). Lines
and shaded areas are guides for the eye. Inset: evolution of the
FWHM resolution ellipsoid in the course of a typical inelastic
scan used in this study. The solid line is the dispersion relation
for the central coherent mode in NTENP at $H=13.8$~T. The dashed
line represents the slope defined by the spin wave
velocity.\label{raw}}
\end{figure}

\begin{figure}
\includegraphics[width=3.2in]{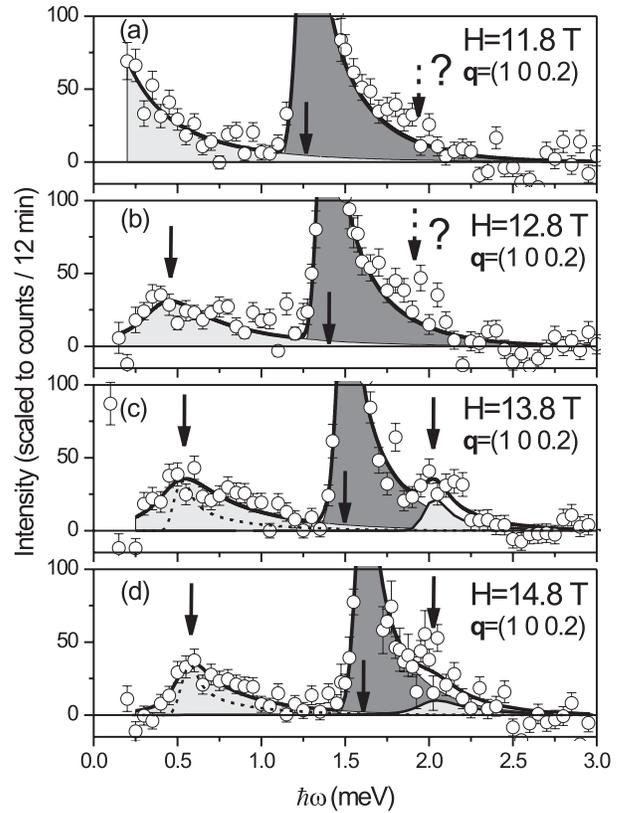}
\caption{Background-subtracted inelastic scans measured at the 1D
AF zone-center $\mathbf{q}_0=(1,0,0.2)$ in NTENP at various fields
(open circles). Solid lines are fits to the data as described in
the text. The shaded areas are the corresponding contributions of
the three terms in the model cross section. The dashed line in
(b)--(d) represent the experimental resolution, shown to emphasize
the intrinsic broadening of the lower-energy peak. Arrows indicate
the three main features of the spectrum. \label{exdata}}
\end{figure}

\begin{figure}
\includegraphics[width=3.2in]{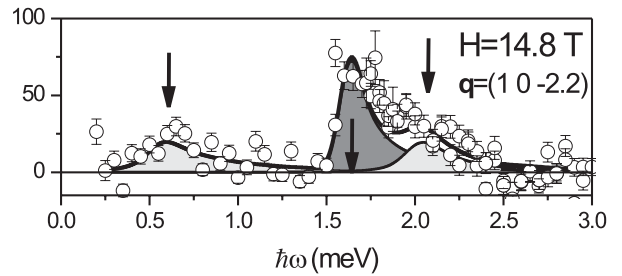}
\caption{Background-subtracted inelastic scan measured at the 1D
AF zone-center $\mathbf{q}_1=(1,0,-2.2)$ in NTENP at $H=14.8$~T
(open circles). Solid lines are fits to the data as described in
the text. The shaded areas are the corresponding contributions of
the three terms in the model cross section.  Arrows indicate the
three main features of the spectrum. \label{exdata2}}
\end{figure}

The data were collected in constant-$q$ scans near the 1D AF
zone-centers $\mathbf{q}_0=(1,0,0.2)$ and
$\mathbf{q}_1=(1,0,-2.2)$. Typical raw data are shown in
Fig.~\ref{raw} (open circles). The background (Fig.~\ref{raw},
solid circles) was measured at $\mathbf{q}=(0.8,0,0.2)$ and
$\mathbf{q}=(1.2,0,0.2)$, and found to be field-independent. A
5-point smoothed average of all background scans was subtracted
from the data at each field. The resulting data sets collected at
several fields are shown in Figs.~\ref{exdata} and~\ref{exdata2}.
They are to be compared with the data shown in Figs.~2 and 4 of
Ref.~\onlinecite{Hagiwara2005}, that were measured on a
considerably less intense neutron source, with a somewhat broader
energy resolution.

The new data for $H=11.8~T \approx H_c$ are consistent with
previous measurements. They clearly show a gap mode at $\hbar
\omega=1.3$~meV and a strong gapless feature at low energies. The
latter corresponds to the excitation branch that went soft at the
critical field. At higher fields, the scattering intensity at low
energies decreases, but does not vanish entirely, as previously
assumed.\cite{Hagiwara2005} Instead, it is transformed into a
broad peak (lower light shaded area in Fig.~\ref{raw}a) that is
positioned below the strong coherent mode (dark shaded area in
Fig.~\ref{raw}a). The low intensity of this new feature explains
why it could not be reliably detected in previous experiments:
only a hint of a low-energy peak is visible in Fig.~4c of
Ref.~\onlinecite{Hagiwara2005}. The new feature persists at higher
fields, and its gap increases, with an intensity maximum shifting
to around 0.6~meV at $H=14.8~T$. At this field it is also observed
at an equivalent but larger and almost orthogonal scattering
vector $\mathbf{q}_1=(1,0,-2.2)$ (Fig.~\ref{exdata2}). Here the
intensity is considerably smaller, consistent with its magnetic
(as opposed to vibrational) origin. It is important to emphasize
that, unlike the sharp and intense lowest-energy mode in NDMAP,
the low-energy feature in NTENP remains very broad and weak in the
entire accessible field range.

At the highest fields, yet another new feature of the spectrum is
revealed at $\hbar \omega \approx 2$~meV energy transfer (upper
light shaded area in Fig.~\ref{raw}a, and Figs.~\ref{exdata}c,d).
The new peak is also detected at $\mathbf{q}_1=(1,0,-2.2)$
(Fig.~\ref{exdata2}). This peak clearly does {\it not} correspond
to the highest-energy member of the triplet seen at low fields.
Indeed, the latter vanishes well below $H_c$, and it energy
extrapolates to much higher values with increasing field
(somewhere in the range 3--3.5~meV at $14.8$~T).  Although at
present we can not irrefutably prove that the new feature does not
originate from spurious scattering, that possibility seems rather
unlikely. Indeed, the peak is observed at several wave vectors,
does not increase at a larger wave vector,  shows a clear field
dependence, and, as will be discussed below, appears to have a
meaningful polarization character. In the discussion below we
shall assume that it is in fact real and magnetic in nature.

\section{Data analysis and discussion}
\begin{figure}
\includegraphics[width=3.2in]{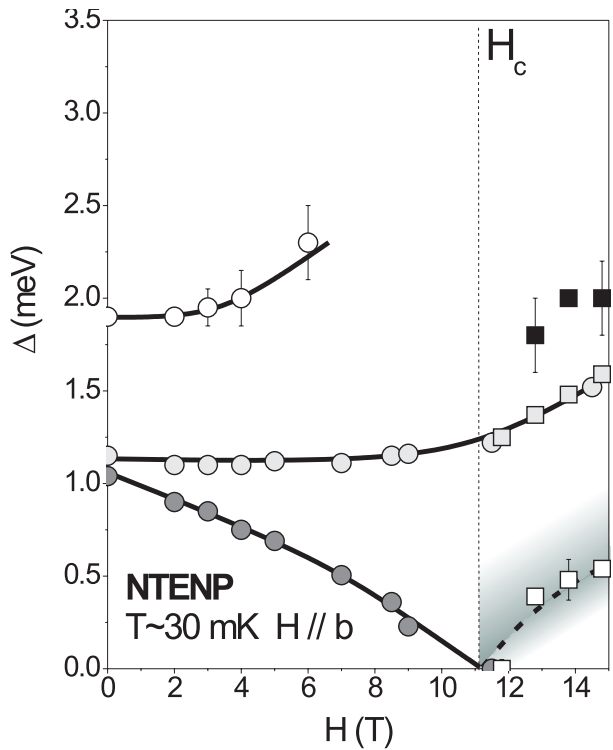}
\caption{Field dependence of excitation energies at the 1D AF
zone-center in NTENP. Open squares are positions of the intensity
maximum in the low-lying excitation continuum (grayed area). Gray
squares are the energies of the coherent central mode. Solid
squares correspond to the ``2~meV'' peak. Circles are from
Ref.~\protect\onlinecite{Hagiwara2005}. \label{vsh}}
\end{figure}

In order to learn more about the newly observed excitations, we
performed fits to the experimental data, taking into account the
effects of instrument resolution. The model cross section used for
these fits included three separate branches, the partial
contribution of each mode written as:
\begin{eqnarray}
\left[\frac{d \sigma}{d \Omega d E'}\right] _\alpha \propto
\frac{1}{\omega_{\mathbf{q},\alpha}}
L\left({\hbar \omega-\hbar \omega_{\mathbf{q},\alpha}},{\Gamma_\alpha}\right),\\
(\hbar
\omega_{\mathbf{q},\alpha})^2=\Delta_\alpha^2+v^2\sin^2(\mathbf{qa}),
\end{eqnarray}
where $(\hbar \omega_{\mathbf{q},\alpha})$ is the dispersion
relation for mode $\alpha$, $L(x,\Gamma)$ is a normalized peak
shape function of width $\Gamma$, taken in the Lorentzian or
Dirac's $\delta$-function (for $\Gamma=0$) form, and $v=8.6$~meV
is the previously determined spin wave
velocity.\cite{Zheludev2004-2,Regnault2004,Hagiwara2005} The model
cross section was numerically convoluted with the resolution
function of the spectrometer that was calculated in the Popovici
approximation.\cite{Popovici75} The intensities and widths of each
mode, as well as the gap energies were refined to best-fit the
experimental data at each field. The results of the fit are shown
in heavy solid lines in Fig.~\ref{exdata}. The grayed areas
represent partial contributions of each mode. The excitation
energies are plotted against the strength of applied field in
Fig.~\ref{vsh} (squares). Here we also show the data from previous
studies (circles).\cite{Hagiwara2005}

The fits consistently yielded a zero intrinsic energy width for
the central coherent mode. However, at $H>H_c$ the lowest-energy
feature is clearly broader than experimental energy resolution.
Its intrinsic energy full width at half height as about
0.30(6)~meV at $H=12.8$~T, and decreases to 0.15(4)~meV at the
highest accessible field of 14.8~T. The broad nature of this
excitation is emphasized by the dashed lines in Fig.~\ref{exdata}
that represent the calculated resolution width. We thus conclude
that at $H>H_c$ the lowest-energy excitations in NTENP are
actually a continuum of states, rather than a single long-lived
coherent mode as, for instance, in NDMAP. Such behavior is fully
consistent with the Lanczos diagonalization results of
Ref.~\onlinecite{Suzuki2005}. Indeed, that numerical study
predicted three coherent modes for NDMAP and only one for NTENP,
but also a low-lying continuum of excitations in the latter
system. In NTENP the observed intensity maximum of the low-energy
spectral feature can be associated with the singularity on the
lower bound of the continuum. This interpretation is consistent
with the observed continuum spectrum being gapped or, at least,
showing a pseudogap behavior.

For the ``2~meV'' excitation, a zero intrinsic width was assumed
for $H=13.8$~T, where it is plainly too weak for a meaningful
peak-shape analysis. However, at $\mathbf{q}_1=(1,0,-2.2)$ at
$H=14.8$~T (Fig.~\ref{exdata2}) the feature is clearly broad
beyond the effect of experimental resolution. At this field its
intrinsic width was determined to be 0.12(6)~meV, comparable to
that of the low-energy continuum.

Additional information on the excitations in the high-filed phase
was obtained in a partial polarization analysis at the highest
accessible filed of $H=14.8$~T. This procedure involved a
comparison of the inelastic intensities measured at
$\mathbf{q}_0=(1,0,0.2)$ (Fig.~\ref{exdata}c) and the equivalent
wave vector $\mathbf{q}_1=(1,0,-2.2)$ (Fig.~\ref{exdata2}). Note
that, due to the large values of the crystallographic angle
$\beta$, these two wave vectors are almost orthogonal to each
other, and form angle of $33^\circ$ and $67^\circ$ with the $a$
(chain) axis, respectively. Neutron spectroscopy probes only the
fluctuations of spin components perpendicular to the scattering
vector. For $\mathbf{q}_0$ the intensity scaling factors for
excitation polarized along the $x$ (parallel to the chains), $y$
(in the horizontal plain, perpendicular to the chains) and $z$
(vertical) axes are 0.30, 0.70 and 1.0, respectively. For
$\mathbf{q}_1$ the corresponding coefficients are 0.85, 0.15 and
1.0. Fits to the experimental data reveal that, corrected for
resolution effects and the Ni$^{2+}$ magnetic form factor, the
intensities of all three components of the spectrum vary between
the two wave vectors. At $\mathbf{q}_1$ the corrected intensities
of the central coherent mode and the low-energy continuum are
decreased by factors of 1.6(1) and 1.27(4), respectively, as
compared to $\mathbf{q}_0$. In contrast, after form factor and
focusing corrections, the ``2meV'' excitation is {\it stronger} at
$\mathbf{q}_1$, by a factor of 2.4(3). This intensity pattern is
consistent with the ``2meV'' excitation being polarized parallel
to the spin chains, and the two lower-energy modes being
transverse in nature.

Regarding the origin of the ``2~meV'' feature, we note that its
peak energy is approximately equal to the sum of the gap in the
main coherent mode and the energy of the intensity maximum of the
continuum. This may indicate that the ``2~meV'' mode is a
composite excitation of these two lower-energy states, similar to
bound magnon states in the classical magnet TMMC.\cite{Heilmann81}
Such excitations near $q=\pi$ are permitted by energy-momentum
conservation laws. Indeed, the intrinsic breaking of translational
symmetry (bond-alternation) in NTENP makes the wave vectors $q=0$
and $q=\pi$ equivalent, allowing excitations of total momentum
$\pi$ composed of a continuum state near $q=0$ and a single-magnon
state at $q=\pi$, or vice versa. A further theoretical study is
required to clarify this point.

\section{Conclusion}
The examples of NDMAP and NENP may lead to a simple-minded
conclusion that the high field phase of anisotropic gapped quantum
spin chains is entirely dominated by long-lived coherent
excitations. The new results on NTENP clearly demonstrate the
opposite: broad spectral continua may, in fact, play a key role in
the dynamics of highly anisotropic quantum spin solids even at low
energies.

\acknowledgments We thank Sei-Ichiro Suga and Takahumi Suzuki for
fruitful discussions. This work was in part supported by the
Molecular Ensemble research program from RIKEN and the
Grant-in-Aid for Scientific Research on Priority Areas from the
Japanese Ministry of Education, Culture, Sports, Science and
Technology. Work at ORNL was sponsored by the Division of
Materials Sciences and Engineering, Office of Basic Energy
Sciences, U.S. Department of Energy, under contract
DE-AC05-00OR22725 with Oak Ridge National Laboratory, managed and
operated by UT-Battelle, LLC.


\begin{thebibliography}{19}
\expandafter\ifx\csname
natexlab\endcsname\relax\def\natexlab#1{#1}\fi
\expandafter\ifx\csname bibnamefont\endcsname\relax
  \def\bibnamefont#1{#1}\fi
\expandafter\ifx\csname bibfnamefont\endcsname\relax
  \def\bibfnamefont#1{#1}\fi
\expandafter\ifx\csname citenamefont\endcsname\relax
  \def\citenamefont#1{#1}\fi
\expandafter\ifx\csname url\endcsname\relax
  \def\url#1{\texttt{#1}}\fi
\expandafter\ifx\csname
urlprefix\endcsname\relax\def\urlprefix{URL }\fi
\providecommand{\bibinfo}[2]{#2}
\providecommand{\eprint}[2][]{\url{#2}}

\bibitem[{\citenamefont{Schulz}(1986)}]{Schulz86}
\bibinfo{author}{\bibfnamefont{H.~J.} \bibnamefont{Schulz}},
  \bibinfo{journal}{Phys. Rev. B} \textbf{\bibinfo{volume}{34}},
  \bibinfo{pages}{6372} (\bibinfo{year}{1986}).

\bibitem[{\citenamefont{Katsumata et~al.}(1989)\citenamefont{Katsumata, Hori,
  T.~Takeuchi, Yamagishi, and Renard}}]{Katsumata89}
\bibinfo{author}{\bibfnamefont{K.}~\bibnamefont{Katsumata}},
  \bibinfo{author}{\bibfnamefont{H.}~\bibnamefont{Hori}},
  \bibinfo{author}{\bibfnamefont{M.~D.} \bibnamefont{T.~Takeuchi}},
  \bibinfo{author}{\bibfnamefont{A.}~\bibnamefont{Yamagishi}},
  \bibnamefont{and} \bibinfo{author}{\bibfnamefont{J.~P.}
  \bibnamefont{Renard}}, \bibinfo{journal}{Phys. Rev. Lett.}
  \textbf{\bibinfo{volume}{63}}, \bibinfo{pages}{86} (\bibinfo{year}{1989}).

\bibitem[{\citenamefont{Affleck}(1990)}]{Affleck90}
\bibinfo{author}{\bibfnamefont{I.}~\bibnamefont{Affleck}},
  \bibinfo{journal}{Phys. Rev. Lett.} \textbf{\bibinfo{volume}{65}},
  \bibinfo{pages}{2477} (\bibinfo{year}{1990}).

\bibitem[{\citenamefont{Takahashi and Sakai}(1991)}]{Takahashi91}
\bibinfo{author}{\bibfnamefont{M.}~\bibnamefont{Takahashi}} \bibnamefont{and}
  \bibinfo{author}{\bibfnamefont{T.}~\bibnamefont{Sakai}}, \bibinfo{journal}{J.
  Phys. Soc. Jpn.} \textbf{\bibinfo{volume}{60}}, \bibinfo{pages}{760}
  (\bibinfo{year}{1991}).

\bibitem[{\citenamefont{Honda et~al.}(1998)\citenamefont{Honda, Asakawa, and
  Katsumata}}]{Honda98}
\bibinfo{author}{\bibfnamefont{Z.}~\bibnamefont{Honda}},
  \bibinfo{author}{\bibfnamefont{H.}~\bibnamefont{Asakawa}}, \bibnamefont{and}
  \bibinfo{author}{\bibfnamefont{K.}~\bibnamefont{Katsumata}},
  \bibinfo{journal}{Phys. Rev. Lett.} \textbf{\bibinfo{volume}{81}},
  \bibinfo{pages}{2566} (\bibinfo{year}{1998}).

\bibitem[{\citenamefont{Chen et~al.}(2001)\citenamefont{Chen, Honda, Zheludev,
  Broholm, Katsumata, and Shapiro}}]{Chen2001}
\bibinfo{author}{\bibfnamefont{Y.}~\bibnamefont{Chen}},
  \bibinfo{author}{\bibfnamefont{Z.}~\bibnamefont{Honda}},
  \bibinfo{author}{\bibfnamefont{A.}~\bibnamefont{Zheludev}},
  \bibinfo{author}{\bibfnamefont{C.}~\bibnamefont{Broholm}},
  \bibinfo{author}{\bibfnamefont{K.}~\bibnamefont{Katsumata}},
  \bibnamefont{and} \bibinfo{author}{\bibfnamefont{S.~M.}
  \bibnamefont{Shapiro}}, \bibinfo{journal}{Phys. Rev. Lett.}
  \textbf{\bibinfo{volume}{86}}, \bibinfo{pages}{1618} (\bibinfo{year}{2001}).

\bibitem[{\citenamefont{Zheludev et~al.}(2002)\citenamefont{Zheludev, Z.~Honda,
  Broholm, and Katsumata}}]{Zheludev2002}
\bibinfo{author}{\bibfnamefont{A.}~\bibnamefont{Zheludev}},
  \bibinfo{author}{\bibfnamefont{Y.~C.} \bibnamefont{Z.~Honda}},
  \bibinfo{author}{\bibfnamefont{C.}~\bibnamefont{Broholm}}, \bibnamefont{and}
  \bibinfo{author}{\bibfnamefont{K.}~\bibnamefont{Katsumata}},
  \bibinfo{journal}{Phys. Rev. Lett.} \textbf{\bibinfo{volume}{88}},
  \bibinfo{pages}{077206} (\bibinfo{year}{2002}).

\bibitem[{\citenamefont{Zheludev et~al.}(2003)\citenamefont{Zheludev, Honda,
  Broholm, Katsumata, Shapiro, Kolezhuk, Park, and Qiu}}]{Zheludev2003}
\bibinfo{author}{\bibfnamefont{A.}~\bibnamefont{Zheludev}},
  \bibinfo{author}{\bibfnamefont{Z.}~\bibnamefont{Honda}},
  \bibinfo{author}{\bibfnamefont{C.}~\bibnamefont{Broholm}},
  \bibinfo{author}{\bibfnamefont{K.}~\bibnamefont{Katsumata}},
  \bibinfo{author}{\bibfnamefont{S.~M.} \bibnamefont{Shapiro}},
  \bibinfo{author}{\bibfnamefont{A.}~\bibnamefont{Kolezhuk}},
  \bibinfo{author}{\bibfnamefont{S.}~\bibnamefont{Park}}, \bibnamefont{and}
  \bibinfo{author}{\bibfnamefont{Y.}~\bibnamefont{Qiu}},
  \bibinfo{journal}{Phys. Rev. B} \textbf{\bibinfo{volume}{68}},
  \bibinfo{pages}{134438} (\bibinfo{year}{2003}).

\bibitem[{Zhe()}]{Zheludev2005rev}
\bibinfo{note}{A. Zheludev, cond-mat/0507534 (2005)}.

\bibitem[{Reg({\natexlab{a}})}]{Regnault}
\bibinfo{note}{L.P. Regnault, I. Zaliznyak, J.P. Renard, and C. Vettier, Phys. Rev. B \textbf{50}, 9174 (1994)}.


\bibitem[{\citenamefont{Narumi et~al.}(2001)\citenamefont{Narumi, Hagiwara,
  Kohno, and Kindo}}]{Narumi2001}
\bibinfo{author}{\bibfnamefont{Y.}~\bibnamefont{Narumi}},
  \bibinfo{author}{\bibfnamefont{M.}~\bibnamefont{Hagiwara}},
  \bibinfo{author}{\bibfnamefont{M.}~\bibnamefont{Kohno}}, \bibnamefont{and}
  \bibinfo{author}{\bibfnamefont{K.}~\bibnamefont{Kindo}},
  \bibinfo{journal}{Phys. Rev. Lett.} \textbf{\bibinfo{volume}{86}},
  \bibinfo{pages}{324} (\bibinfo{year}{2001}).

\bibitem[{\citenamefont{Zheludev et~al.}(2004)\citenamefont{Zheludev, Masuda,
  Sales, Mandrus, Papenbrock, Barnes, and Park}}]{Zheludev2004-2}
\bibinfo{author}{\bibfnamefont{A.}~\bibnamefont{Zheludev}},
  \bibinfo{author}{\bibfnamefont{T.}~\bibnamefont{Masuda}},
  \bibinfo{author}{\bibfnamefont{B.}~\bibnamefont{Sales}},
  \bibinfo{author}{\bibfnamefont{D.}~\bibnamefont{Mandrus}},
  \bibinfo{author}{\bibfnamefont{T.}~\bibnamefont{Papenbrock}},
  \bibinfo{author}{\bibfnamefont{T.}~\bibnamefont{Barnes}}, \bibnamefont{and}
  \bibinfo{author}{\bibfnamefont{S.}~\bibnamefont{Park}},
  \bibinfo{journal}{Phys. Rev. B} \textbf{\bibinfo{volume}{69}},
  \bibinfo{pages}{144417} (\bibinfo{year}{2004}).

\bibitem[{Reg({\natexlab{b}})}]{Regnault2004}
\bibinfo{note}{L. P. Regnault, M. Hagiwara, N. Metoki, Y. Koike, K. Kakurai,
and A. Stunault, Physica B \textbf{66}, 350B (2004)}.

\bibitem[{\citenamefont{Hagiwara et~al.}(2005)\citenamefont{Hagiwara, Regnault,
  Zheludev, Stunault, Metoki, Suzuki, Suga, Kakurai, Koike, Vorderwisch
  et~al.}}]{Hagiwara2005}
\bibinfo{author}{\bibfnamefont{M.}~\bibnamefont{Hagiwara}},
  \bibinfo{author}{\bibfnamefont{L.~P.} \bibnamefont{Regnault}},
  \bibinfo{author}{\bibfnamefont{A.}~\bibnamefont{Zheludev}},
  \bibinfo{author}{\bibfnamefont{A.}~\bibnamefont{Stunault}},
  \bibinfo{author}{\bibfnamefont{N.}~\bibnamefont{Metoki}},
  \bibinfo{author}{\bibfnamefont{T.}~\bibnamefont{Suzuki}},
  \bibinfo{author}{\bibfnamefont{S.}~\bibnamefont{Suga}},
  \bibinfo{author}{\bibfnamefont{K.}~\bibnamefont{Kakurai}},
  \bibinfo{author}{\bibfnamefont{Y.}~\bibnamefont{Koike}},
  \bibinfo{author}{\bibfnamefont{P.}~\bibnamefont{Vorderwisch}},
  \bibnamefont{et~al.}, \bibinfo{journal}{Phys. Rev. Lett.}
  \textbf{\bibinfo{volume}{94}}, \bibinfo{pages}{177202}
  (\bibinfo{year}{2005}).

\bibitem[{\citenamefont{Suzuki and Suga}(2005)}]{Suzuki2005}
\bibinfo{author}{\bibfnamefont{T.}~\bibnamefont{Suzuki}} \bibnamefont{and}
  \bibinfo{author}{\bibfnamefont{S.}~\bibnamefont{Suga}},
  \bibinfo{journal}{Phys. Rev. B} \textbf{\bibinfo{volume}{72}},
  \bibinfo{pages}{014434} (\bibinfo{year}{2005}).

\bibitem[{\citenamefont{Kolezhuk and Sachdev}()}]{Kolezhuk2005}
\bibinfo{author}{\bibfnamefont{A.}~\bibnamefont{Kolezhuk}} \bibnamefont{and}
  \bibinfo{author}{\bibfnamefont{S.}~\bibnamefont{Sachdev}},
  \bibinfo{note}{cond-mat/0511353.}

\bibitem[{\citenamefont{Zhitomirsky}()}]{Zhitomirsky2005}
\bibinfo{author}{\bibfnamefont{M.~E.} \bibnamefont{Zhitomirsky}},
  \bibinfo{note}{cond-mat/0601405.}

\bibitem[{\citenamefont{Popovici}(1975)}]{Popovici75}
\bibinfo{author}{\bibfnamefont{M.}~\bibnamefont{Popovici}},
  \bibinfo{journal}{Acta Cryst.} \textbf{\bibinfo{volume}{A31}},
  \bibinfo{pages}{507} (\bibinfo{year}{1975}).

\bibitem[{\citenamefont{Heilmann et~al.}(1981)\citenamefont{Heilmann, Kjems,
  Endoh, Reiter, Shirane, and Birgeneau}}]{Heilmann81}
\bibinfo{author}{\bibfnamefont{I.}~\bibnamefont{Heilmann}},
  \bibinfo{author}{\bibfnamefont{J.}~\bibnamefont{Kjems}},
  \bibinfo{author}{\bibfnamefont{Y.}~\bibnamefont{Endoh}},
  \bibinfo{author}{\bibfnamefont{G.}~\bibnamefont{Reiter}},
  \bibinfo{author}{\bibfnamefont{G.}~\bibnamefont{Shirane}}, \bibnamefont{and}
  \bibinfo{author}{\bibfnamefont{R.}~\bibnamefont{Birgeneau}},
  \bibinfo{journal}{Phys. Rev. B} \textbf{\bibinfo{volume}{24}},
  \bibinfo{pages}{3939} (\bibinfo{year}{1981}).

\end{thebibliography}

\end{document}